\documentclass[preprint,12pt]{elsarticle}

\newcommand{\aneupy}{\textbf{AneuPy}}
\usepackage{graphicx}
\usepackage{amssymb}
\usepackage{amsmath}
\usepackage{algpseudocode}
\usepackage{algorithm}
\usepackage{multirow}
\usepackage{bm}
\usepackage{color,soul}

\usepackage{listings}
\usepackage{xcolor}
\usepackage[margin=2.5cm]{geometry}
\usepackage[%
    colorlinks=true,%
    hyperindex,%
    plainpages=false,%
    bookmarksopen,%
    bookmarksnumbered]{hyperref}

\newcommand{\software}[1]{\texttt{#1}} 
\newcommand{\fileformat}[1]{\texttt{.#1}} 

\makeatletter
\AtBeginDocument{\def\@citecolor{blue}}
\makeatletter
\def\ps@pprintTitle{%
 \let\@oddhead\@empty
 \let\@evenhead\@empty
 \def\@oddfoot{}%
 \let\@evenfoot\@oddfoot}
\makeatother

\begin{document}

\begin{frontmatter}


\title{\textbf{AneuPy}: An open source Python tool for creating simulation-ready geometries of abdominal aortic aneurysms}



\author[label1]{Mario de Lucio$^*$\corref{Corresponding author:mdeluci@purdue.edu}}
\author[ad2]{Jacobo Díaz}
\author[ad2]{Alberto de Castro}
\author[ad2]{Luis E. Romera}
\cortext[cor1]{Corresponding author: mdeluci@purdue.edu}

\address[label1]{School of Mechanical Engineering, Purdue University, 585 Purdue Mall, West Lafayette, Indiana 47907, USA}
\address[ad2]{Universidade da Coruña, Center for Technological Innovation in Construction and Civil Engineering (CITEEC), Campus de Elviña, 15071, A Coruña, Spain}

\begin{abstract}
Abdominal aortic aneurysms (AAAs) are localized dilatations of the abdominal aorta that can lead to life-threatening rupture if left untreated. AAAs primarily affect older individuals, with high mortality rates following rupture, so early diagnosis and risk assessment are critical. The geometrical characteristics of an AAA, such as its maximum diameter, asymmetry, and wall thickness, are extremely significant in biomechanical models for the assessment of rupture risk. Despite the growing use of computational modeling for AAA investigation, there is a notable gap in accessible, open-source software capable of generating simulation-ready geometries for biomechanical and hemodynamic simulations. To address this gap, we introduce \textbf{AneuPy}, an open-source Python-based tool designed to create both idealized and patient-specific AAA geometric models. \textbf{AneuPy} is a fast and automated approach for generating aneurysm geometries from minimal input data, allowing for extensive parameter customization. By automating the creation of simulation-ready geometries for finite element analysis (FEA), computational fluid dynamics (CFD), or fluid-structure interaction (FSI) models, \textbf{AneuPy} can facilitate research in AAA and improve patient-specific risk prediction.

\end{abstract}


\end{frontmatter}
\section{Motivation and significance}
\label{intro}
Abdominal aortic aneurysms (AAAs) are localized dilatations of the abdominal aorta, defined by an expansion of the aorta to a diameter greater than 1.5 times its normal diameter \cite{AAA2}. While AAAs can grow asymptomatically, their potential for rupture, coupled with high associated mortality, makes them a serious global health concern. Worldwide, AAAs result in more than 175,000 deaths each year, representing about 1\% of mortality in men aged over 65 years old \cite{AAAs1}. Given this risk, early diagnosis and accurate risk assessment are vital for determining appropriate surgical intervention.

The geometry of an AAA is a critical focus in biomechanical research, with factors such as aneurysm size, shape, and wall thickness directly influencing hemodynamic forces and wall stress distributions. Computational modeling has demonstrated that geometric features like maximum diameter, aspect ratio, and asymmetry significantly influence rupture risk predictions \cite{Doyle2007,Polzer2013}. These geometric properties are known to result in more accurate patient-specific predictions, improving monitoring and treatment planning for AAAs.

Beyond geometry, accurate assignment of material properties is crucial for AAA modeling, since wall thickness, fiber orientation, and local stiffness impact stress distributions and rupture risk predictions. Patient-specific geometries derived from medical imaging further enhance model accuracy, allowing for more realistic simulations of aneurysm morphology and behavior \cite{DELUCIO2023105602}. It has been shown that high-fidelity computational approaches that take advantage of patient-specific data enhance simulations and predictive capabilities \cite{deLucio3,DELUCIO2024124446}.

Several software tools currently support the generation and analysis of AAA geometries. Commercial software like \software{MIMICS Innovation Suite} by Materialise offers robust 3D segmentation, modeling, and visualization of medical imaging data, commonly used for patient-specific simulations \cite{MIMICS}.
\software{3D Slicer}, an open-source software platform, is widely utilized for image processing, 3D visualization, and segmentation, accommodating various computational approaches for AAA modeling \cite{3DSlicer}. \software{Synopsys} is a commercial software that has powerful capabilities for advanced geometric modeling and simulation, though primarily focused on industrial application \cite{SynopsysSimpleware}. The \software{Vascular Modeling Toolkit} (VMTK) is an open-source collection of tools committed to the generation and manipulation of vascular geometries, including mesh generation, analysis, and fluid dynamics simulations \cite{VMT}. \software{ITK SNAP} is another popular open source tool, primarily utilized for image segmentation and 3D visualization, but without direct coupling to computational modeling \cite{ITKSnap}. 

Despite all these initiatives, these tools have a tendency to be specialized in certain aspects of geometry generation, such as segmentation or meshing, and are maybe not so well optimized for the generation of simulation-ready geometries with pre-defined biomechanical properties. The need for an streamlined but flexible tool for AAA geometry creation that is simple to use but can create idealized as well as patient-specific models has been the motivating factor behind the development of \textbf{AneuPy}.

\section{Software description}
\textbf{AneuPy} utilizes the Python interface of \software{SALOME} \cite{2015salome}, an open source platform for pre- and post-processing for numerical simulations, to automate and streamline the creation of idealized geometries of abdominal aortic aneurysms. Specifically, \textbf{AneuPy} employs \software{SALOME}'s \texttt{Geometry} module, which is accessed through the \texttt{GEOM} Python package. Through this interface, we leverage \software{SALOME}'s robust CAD features to define domains, create cross-sectional geometries, interpolate shells, and obtain the final solid geometries. \textbf{AneuPy} is currently released under the GNU General Public License (GPLv3) \cite{gnu_gpl3}. The codebase is publicly accessible on GitHub \href{https://github.com/mdeluci/AneuPy}{https://github.com/mdeluci/AneuPy}.

\subsection{Architecture}
The cornerstone of \aneupy's architecture resides in the \texttt{Geometry.py} module, which contains all the classes necessary for the geometric modeling process. This ensures a cohesive workflow, making the software easier to navigate, and adaptable. The main components within \texttt{Geometry.py} are:
\begin{description}
    \item[Domain Class:]  Serves as the backbone of \aneupy's architecture. It is responsible for initializing \software{SALOME}'s geometry module and importing all the necessary libraries by calling \texttt{geomBuilder.New()}. Its hierarchical building approach allows for the creation of complex 3D geometric forms (shells, solids) from simple 2D components (sections). It also contains methods to export the geometric models in various formats such as IGES, VTK, BREP or STEP. 
    \item[Section Class:] Defines a cross section. Cross sections are defined in the XY plane and then transformed to the local coordinate system (LCS). Providing the origin is mandatory, and the default LCS is the global coordinate system (GCS). \textbf{AneuPy} uses circles as the geometric entity appended to the cross section. This class is also responsible for adding a circle to the section using a specified center, normal vector and radius.
    \item[Shell Class:] Creates a shell or surface model from multiple sectional geometries using non-Uniform rational B-splines (NURBS). The class allows detailed specification of the NURBS parameters, including minimum and maximum degrees (\texttt{theMinDeg}, \texttt{theMaxDeg}), tolerances for 2D and 3D operations (\texttt{theTol2D}, \texttt{theTol3D}), number of iterations (\texttt{theNbIter}), and sewing precision.
    \item[Solid Class:] Encapsulates the operations related to the final solid geometries within the \software{SALOME} platform. This class is responsible for transitioning from surface geometries to 3D solid geometries. \textbf{AneuPy} uses this class to create solids in two different ways: i) using \texttt{MakeSolid()}, which creates a solid bounded by a closed shell, or ii) \texttt{MakeCut()}, which performs a cut boolean operation on two given shapes.
\end{description}

\subsection{Workflow and extensibility}
\begin{figure}[h!]
    \centering
    \includegraphics[width=0.99\columnwidth]{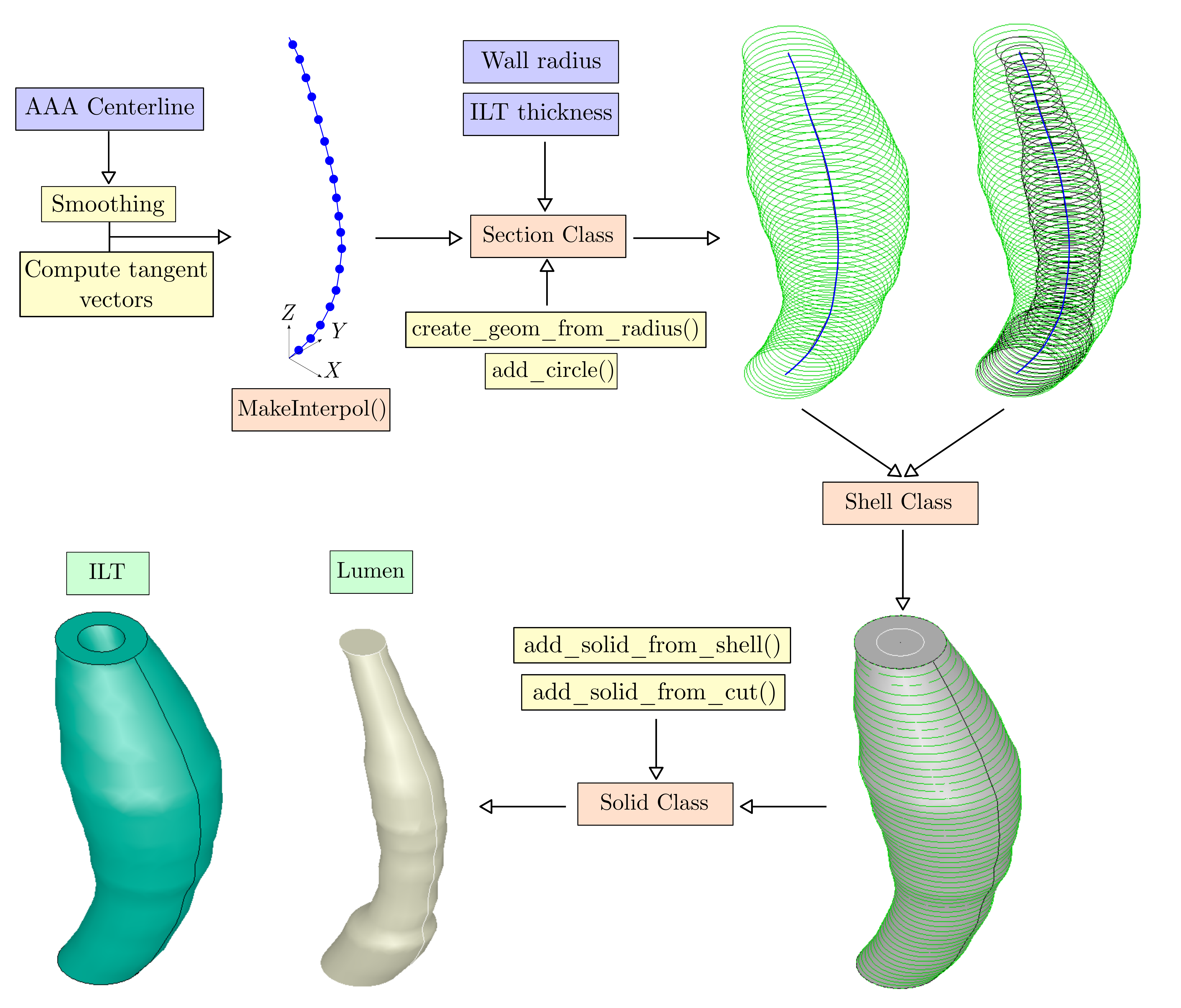}
    \caption{Typical workflow for generating AAA Geometries using \textbf{AneuPy's} \texttt{Patient\_specific.py} module. The process begins with the importation of the aneurysm's centerline as XYZ coordinates. This raw centerline data is then smoothed using cubic B-splines to reduce noise. Points of interest (POIs) are extracted from the smoothed centerline for detailed analysis. Using \texttt{SALOME}'s \texttt{MakeInterpol()} function, a B-spline curve is created. At each POI, a cross-sectional plane is established using the Section class. Subsequently, local radius and ILT thickness are interpolated from the data, and a circle is constructed on each cross-sectional plane. NURBS are interpolated over these circle sections to create a continuous shell along the aneurysm with the \textbf{Shell class}. This shell is then converted into a solid model using \texttt{add\_solid\_from\_shell()}. Finally, boolean operations are performed with \texttt{add\_solid\_from\_cut()} to subtract the ILT volume from the AAA model, generating the final geometries.}
    \label{Algorithm}
\end{figure}

A typical \textbf{AneuPy} workflow is illustrated in Fig.~\ref{Algorithm}. The process consists of the following steps:  

\begin{enumerate}
    \item \textbf{Import Centerline Data} – The aneurysm’s centerline is imported as a series of XYZ coordinates.
    \item \textbf{Smoothing} – A cubic B-spline is applied to reduce noise in the centerline data.
    \item \textbf{Point of Interest (POI) Extraction} – Key locations along the centerline are identified for further processing.
    \item \textbf{Curve Interpolation} – A smooth B-spline curve is generated using \texttt{SALOME}’s \texttt{Make\allowbreak Interpol()} function.
    \item \textbf{Cross-Section Generation} – At each POI, the \texttt{Section} class computes the tangent vector and creates a cross-sectional plane.
    \item \textbf{Radius and ILT Thickness Interpolation} – Local radius and intraluminal thrombus (ILT) thickness are interpolated, and a circle is constructed on each cross-sectional plane.
    \item \textbf{Shell Creation} – NURBS interpolation is applied to the circle sections, generating a continuous shell along the aneurysm using the \texttt{Shell} class.
    \item \textbf{Solid Model Creation} – The shell is converted into a solid model using the \texttt{add\allowbreak\_solid\allowbreak\_from\allowbreak\_shell()} function.
    \item \textbf{Boolean Operations} – The ILT volume is subtracted from the AAA model using \texttt{add\_solid\_from\_cut()}, resulting in the final geometry.
\end{enumerate}

\subsection{Software functionalities}\
\textbf{Parametric creation of AAA geometries:}
\textbf{AneuPy} enables the creation of AAA geometries from parameters given by the user, allowing reproducibility and customization. Users can choose to specify the number, location, and size of the cross sections that define the geometry of the aneurysm. The software provides flexibility for aneurysm morphology, including fusiform and saccular aneurysms.

\textbf{Definition of cross-sections by automatons:} The \texttt{Section} class provides an easy-to-use method for the definition of cross sections by circular profiles. Cross sections may be translated along a reference centerline whose spatial distribution may be controlled by users. The software performs the transformation into the local system of coordinates automatically and appropriately converts the direction of the normal vector to maintain geometrical consistency.

\textbf{NURBS-Based surface reconstruction:} The \texttt{Shell} class employs NURBS to construct smooth aneurysm surfaces from the given cross sections. The algorithm for generating the surface has the option to smooth the geometry by adjusting the degree and tolerance parameters, delivering a high-quality surface continuity for accurate numerical simulations.

\textbf{Solid model generation:} Using the \texttt{Solid} class, \textbf{AneuPy} can generate fully closed solid geometries from surface models. There are two general methods to generate solids using the software: (i) \texttt{MakeSolid()}, which forms a solid by closing a shell, and (ii) \texttt{MakeCut()}, which uses Boolean operations to form intricate shapes. Such solid models can be exported directly for meshing and further simulation.

\textbf{Exporting geometries in multiple formats:}
\textbf{AneuPy} can export the generated geometries in a number of standard formats including IGES, VTK, BREP, and STEP. These formats are supported by a number of simulation packages and provide flexibility for integration into multiple computational workflows.

\textbf{Automation and scripting interface:} \textbf{AneuPy} is scriptable, and one can automate geometry generation without any human intervention. The package is compatible in complex simulation pipelines and therefore is a valuable tool for researchers and engineers involved in patient-specific or idealized aneurysm modeling.

\section{Illustrative examples}
To demonstrate the capabilities of \textbf{AneuPy}, we present two illustrative examples: the generation of idealized and patient-specific geometries.  

\subsection{Idealized aneurysm geometries}  
The first set of examples presented are idealized AAA geometries, which have been extensively used in several computational studies \cite{Scotti2005,Vorp1998,Raghavan2000,deLucio}. These geometries can be generated using two scripts: \texttt{Idealized\_manual.py} and \texttt{Idealized\_automatic.py}. 

\begin{figure}[h!]
    \centering
    \includegraphics[width=0.8\columnwidth]{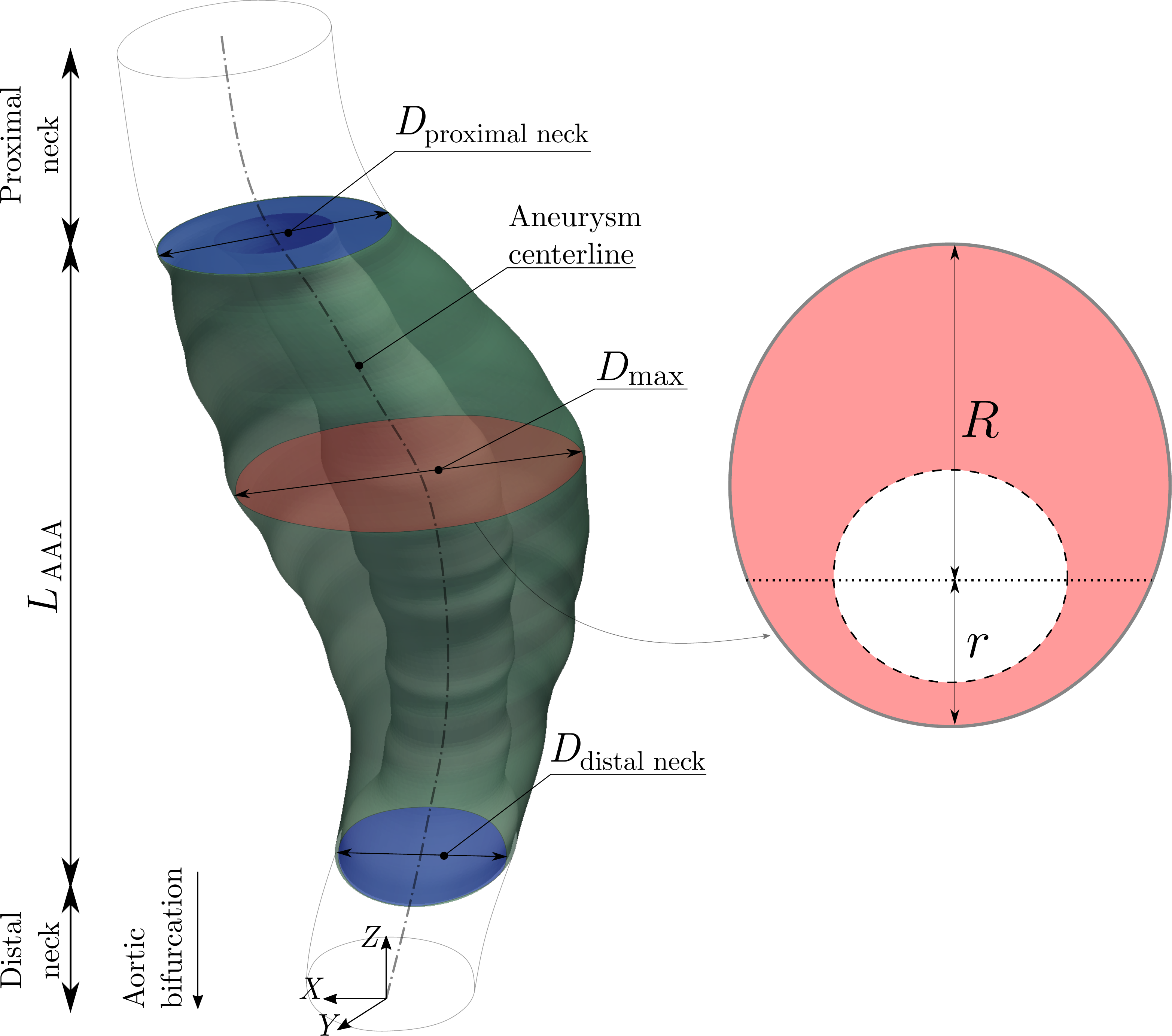}
    \caption{Parameterization of an abdominal aortic aneurysm. Here, $L_{\text{AAA}}$ represents the aneurysm length, $D_{\text{max}}$ is the maximum diameter of the aneurysm, and $D_{\text{proximal neck}}$ and $D_{\text{distal neck}}$ are the diameters of the non-dilated aorta at the proximal and distal necks, respectively. On the right, we show a schematic illustration of the midsection at the location of the maximum diameter, where $r$ and $R$ are the radii measured from the center of the undilated portion to the posterior and anterior walls, respectively.}
    \label{Parameterization}
\end{figure}

Fig.~\ref{Parameterization} illustrates how an AAA can be parameterized using key morphological descriptors. Here, \( D_{\text{max}} \) represents the maximum aneurysm diameter, \( L_{\text{AAA}} \) denotes the aneurysm length, and \( D_{\text{proximal neck}} \) and \( D_{\text{distal neck}} \) correspond to the diameters of the non-dilated aorta at the proximal and distal necks, respectively; $r$ and $R$ are the radii measured at the midsection of the AAA cavity from the longitudinal $Z$-axis to the posterior and anterior walls, respectively. These parameters serve as the foundation for generating geometries in \textbf{AneuPy} and allow for systematic control over aneurysm shape.

To further characterize the aneurysm morphology, we define three dimensionless shape descriptors \cite{Salman,Kleinstreuer2006}:
\begin{equation}
    \beta = \frac{r}{R}, \quad \gamma = \frac{D_{\text{max}}}{L_{\text{AAA}}}, \quad \chi = \frac{D_{\text{max}}}{D_{\text{proximal neck}}}.
\end{equation}
Here, \( \beta \) quantifies the asymmetry of the aneurysm cross-section, \( \gamma \) defines the aspect ratio of the dilation, and \( \chi \) describes the relative expansion compared to the proximal neck diameter. These non-dimensional parameters facilitate comparison across aneurysms of different sizes and have been widely used in computational studies to assess rupture risk \cite{Vorp2005,Fillinger2003}.

\begin{figure}[h!]
    \centering
    \includegraphics[width=0.999\columnwidth]{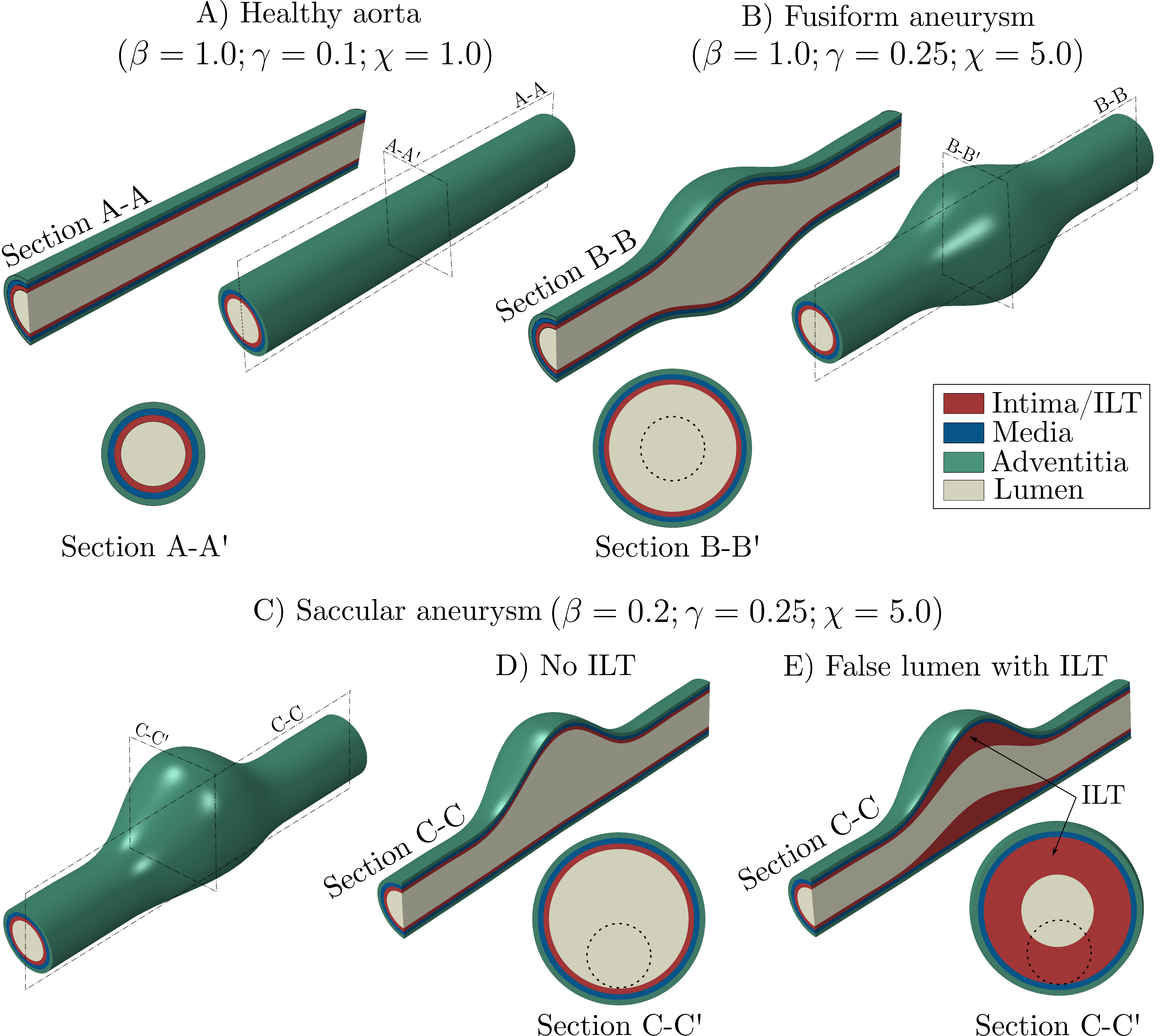}
    \caption{Idealized geometries generated with \textbf{AneuPy}'s \texttt{Idealized\_automatic.py} module. (A) Healthy aorta. (B) Fusiform aneurysm. (C) Saccular aneurysm. (D) Cross sections of a saccular aneurysm without Intra Luminal Thrombus (ILT). (D) Cross sections of a saccular aneurysm with ILT, where the false and true lumen are clearly differentiated. The dotted lines superimposed on the circular cross sections indicate the non-dilated lumen.}
    \label{idealized}
\end{figure}

These parameterized geometries serve as the input for the \textbf{AneuPy} scripts that automate aneurysm shape generation. The \texttt{Idealized\_manual.py} script allows users to manually define the placement of cross-sections along the aneurysm centerline, offering fine control over shape variations. Below is an example of how \texttt{Idealized\_manual.py} works:
\begin{lstlisting}[language=Python, caption={Python script for generating an idealized aneurysm geometry in AneuPy. Only key sections are shown for reference.}, label={lst:aneupy_generation}]
import Geometry
aneupy = Geometry

import salome
salome.salome_init()

d = aneupy.Domain()

# Define cross-sections
d.add_section(name='a1', origin=[0., 0., 0.])
d.add_section(name='a2a', origin=[0., 0., 30.])
d.add_section(name='a3', origin=[0., 0., 50.])
d.add_section(name='a5', origin=[0., 0., 100.])

# Define radii for each section
d.sections['a1'].add_circle(radius=5.)
d.sections['a2a'].add_circle(radius=7.)
d.sections['a3'].add_circle(radius=12.5)
d.sections['a5'].add_circle(radius=5.)

# Generate outer shell of aneurysm
d.add_shell(name='aneurysm_outer', 
            sections=['a1', 'a2a', 'a3', 'a5'],
            minBSplineDegree=10, maxBSplineDegree=20, approximation=True)

# Create solid aneurysm structure from shell
d.add_solid_from_shell(name='aneurysm_outer', shell='aneurysm_outer')

# Save solid as IGES
d.export_iges(solid='aneurysm_outer', file='aneurysm_outer.iges')

\end{lstlisting}
On the other hand, the \texttt{Idealized\_automatic.py} script generates aneurysms from parameters of shape specified by the user, simplifying the process of generating computational models. Fig.~\ref{idealized} demonstrates the efficiency of the \texttt{Idealized\_automatic.py} module for generating idealized aneurysm geometries with various values of the aspect ratios $\chi$, $\gamma$ and asymmetry parameter $\beta$. This module offers the capability to produce a wide range of geometries, from typical aortas to different types of aneurysms, saccular and fusiform, with or without ILT. The geometries are ready-to-mesh and ready for CFD, FEM simulations, and FSI analysis. In addition, the generated geometry comprises all the layers necessary to define the mechanical properties of the aortic wall, including the intima, media, and adventitia. The structured representation facilitates the tagging of material properties to each layer and ease the application of boundary conditions for computational analyses.

The user has two options for running \texttt{Idealized\_automatic.py}: 
\begin{itemize}    
    \item Manual input of the parameters. The script can be executed directly from the command line with user-specified parameters:
    \begin{lstlisting}[language=bash]
    ./Run_Idealized_Automatic.sh --length 120 --radius_nondilated 3 --radius_dilated 8 --wall_thickness_intima 0.5 --wall_thickness_media 0.3 --wall_thickness_adventitia 0.7 --wall_thickness_ILT 2 --x_shift 1.5 --y_shift 2.0
    \end{lstlisting}
    \item Using the JSON configuration file \texttt{Params\allowbreak\_Idealized\allowbreak\_Automatic.\allowbreak json}. Instead of manually specifying parameters, the user can define them in the JSON configuration file and run the code as:
    \begin{lstlisting}[language=bash]
    ./Run_Idealized_Automatic.sh --config_file ./Params_Idealized_Automatic
    \end{lstlisting}
\end{itemize}

The predefined parameters include:  
\begin{itemize}  
    \item \textbf{Length} – Total length of the aneurysm (\( L_{\text{AAA}} \)).  
    \item \textbf{Radius\_nondilated} – Non-dilated radius of the aneurysm (\( D_{\text{proximal neck}} \) and \( D_{\text{distal neck}} \)).  
    \item \textbf{Radius\_dilated} – Radius of the aneurysm sac ($D_{\text{max}}$).  
    \item \textbf{Wall\_thickness\_intima} – Wall thickness of the intima layer.  
    \item \textbf{Wall\_thickness\_media} – Wall thickness of the media layer.  
    \item \textbf{Wall\_thickness\_adventitia} – Wall thickness of the adventitia layer.  
    \item \textbf{Wall\_thickness\_ILT} – Wall thickness of the intraluminal thrombus (ILT).  
    \item \textbf{x\_shift} – Asymmetry of the AAA sac in the X-direction ($r$ and $R$).  
    \item \textbf{y\_shift} – Asymmetry of the AAA sac in the Y-direction ($r$ and $R$).  
\end{itemize}  

These parameters enable the creation of a wide range of aneurysm geometries, from healthy to various aneurysm types, which can be used for further analysis or computational studies. This flexibility is essential to understand the biomechanical behavior of AAAs in different scenarios. Examples of FEM simulations using these geometries can be found in \cite{deLucio}.

\subsection{Patient-Specific Aneurysm Models}  
The \texttt{Patient\_specific.py} module facilitates the generation of curved AAA geometries using imaging-derived patient-specific data. The module needs as inputs the centerline (curved), wall area versus length, and lumen area versus length. These are required to construct a more realistic model of the aneurysm, according to the data of each patient. The user only needs to run the provided \texttt{Run\_Patient\_Specific.sh} script after the text files containing the data in the two-column space-separated format are provided. Once the input data is provided, the user can simply run the shell scripts, and the patient-specific geometries will be automatically created based on the provided data. The user can choose to align the sections in the direction of the $Z$-direction, which coincides with the aneurysm length, or in the direction of the tangent vector at each POI along the centerline. The sample data are in \textbf{AneuPy}\texttt{/test/data}.

\begin{figure}[h!]
    \centering
    \includegraphics[width=0.999\columnwidth]{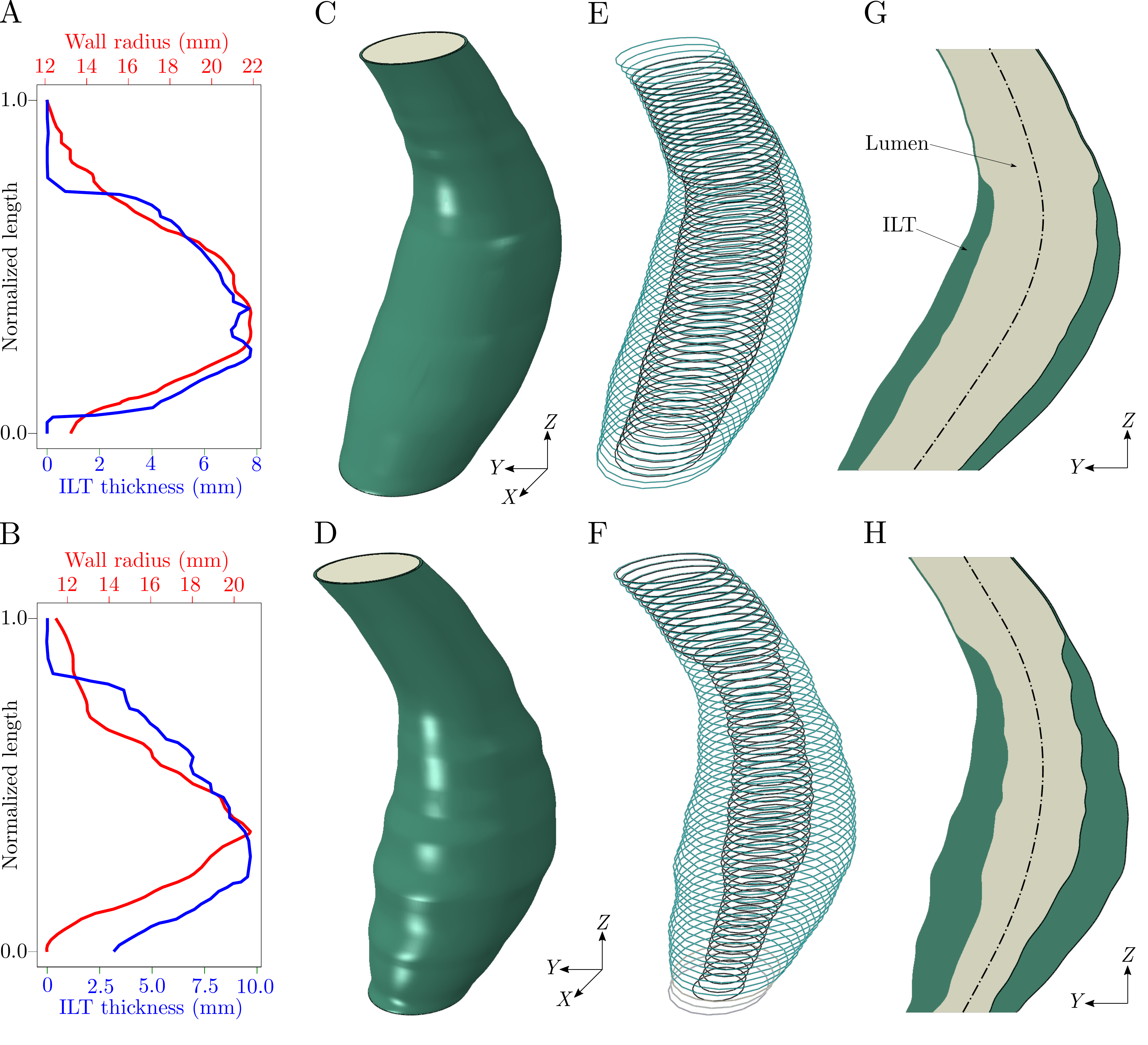}
    \caption{Patient-specific geometries generated with \textbf{AneuPy}'s \texttt{Patient\_specific.py} module. (A-B) Wall radius and ILT thickness vs. normalized length taken from \cite{kontopodis2013changes,kontopodis2020aneurysm}. (C-D) 3D view of generated aneurysms. (E-F) 3D view of cross sections aligned along centerline used to interpolate the outer surfaces. (G-H) Vertical cross section of the aneurysms, where the lumen and the ILT are clearly differentiated. We also show the centerline with a dotted line.}
    \label{PE}
\end{figure}

Wall area vs. length and lumen area vs. length examples employed in the geometry generation process are shown in Fig. \ref{PE}A-B. They are patient-specific and based on data in \cite{kontopodis2013changes,kontopodis2020aneurysm,Gao2020}. Figs. \ref{PE}C-H show two examples of aneurysm geometries, with (C-D) showing the 3D view of the constructed aneurysms. (E-F) show the 3D view of cross sections that are aligned along the centerline and are used to interpolate the outer aneurysm surfaces. Finally, (G-H) show vertical cross sections of the aneurysms, where the lumen and ILT are easily differentiated.

\begin{figure}[h!]
    \centering
    \includegraphics[width=0.99\columnwidth]{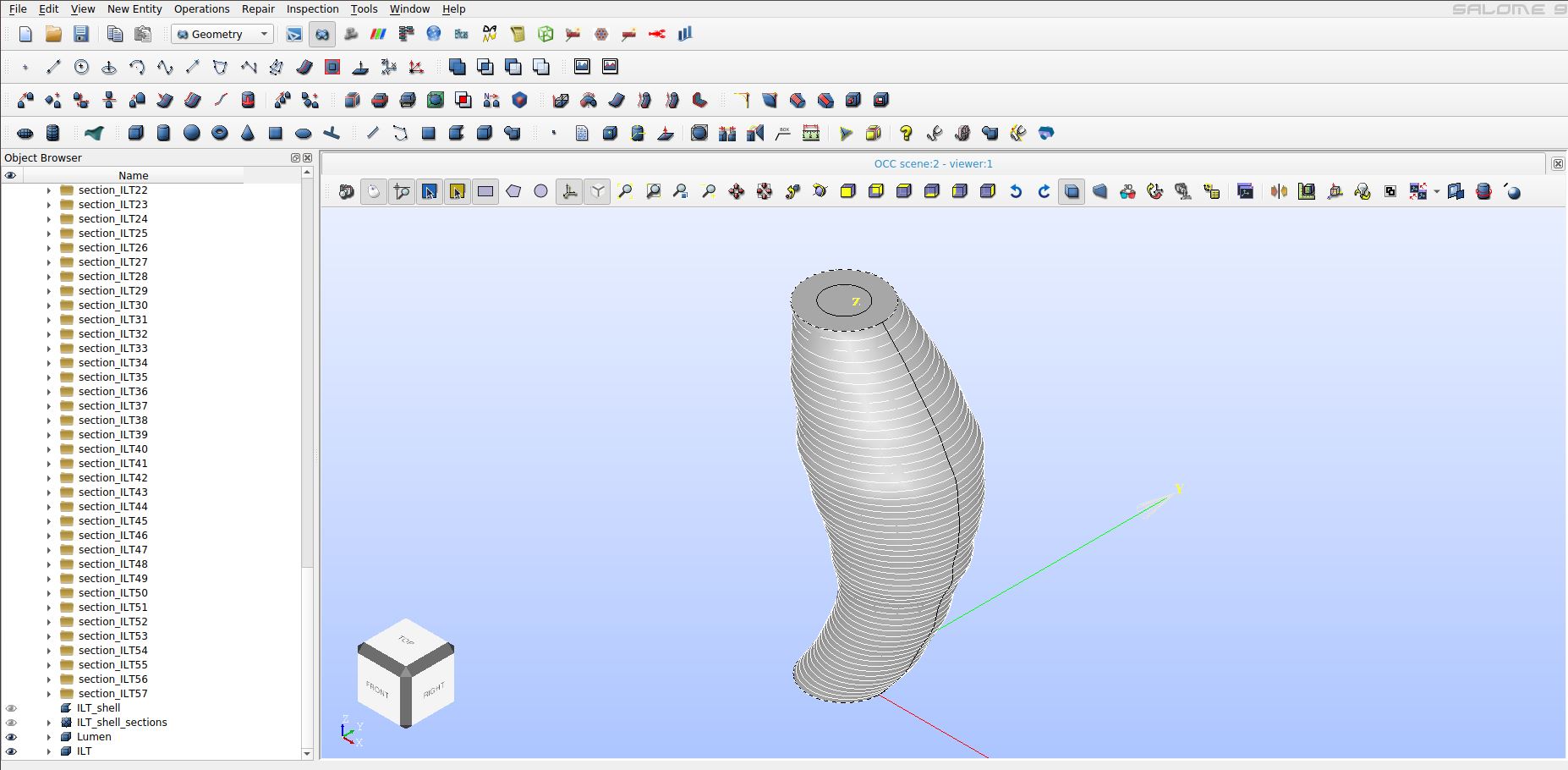}
    \caption{Screenshot of \software{SALOME}'s \texttt{Geometry} module user-friendly interface, where users can open the \fileformat{hdf} study files to inspect the generated geometric entities. This feature allows users to visually verify the created geometries before proceeding with further processing or simulations.}
    \label{salome_interface}
\end{figure}

All codes also generate a \software{SALOME} \fileformat{hdf} study file, which users can open in \software{SALOME}'s \texttt{Geometry} module through its user-friendly interface to inspect the generated geometric entities (see Fig.~\ref{salome_interface}). Although not required, this feature provides a convenient way for users to visualize and verify their geometries during the modeling process.

\section{Conclusions}
\label{conclusions}
In this paper, we presented \textbf{AneuPy}, an open-source Python-based software tool with the goal of simplifying the simulation-ready geometry generation of abdominal aortic aneurysms through automation. The tool automates the geometry generation process using the \software{SALOME} platform to allow users to create idealized as well as patient-specific AAA models for computational simulations. \textbf{AneuPy} is capable of generating geometries ranging from healthy aortas to aneurysms of any shape, including important details such as the aortic wall layers (intima, media, adventitia) and intraluminal thrombus. It also allows for the generation of realistic AAA geometries using imaging-derived patient-specific data, which facilitates improved representation of individual cases in biomechanical studies.

However, there are certain limitations to the software in its current form. Firstly, in the geometry generation, the cross-sections are limited to circular shapes only. Future implementations will incorporate a capability to support higher-order, complex, and irregular geometries. Secondly, the current shell generation code will find it challenging to process highly curved centerlines, e.g., in an aortic arch. This limitation will be addressed in future updates. Lastly, the meshing capabilities are not yet integrated in \textbf{AneuPy}, although they are planned to be included in a future release to further simplify the simulation process.

Despite these limitations, \textbf{AneuPy} presents a robust and versatile software package to generate high-fidelity AAA geometries, allowing a more efficient and streamlined analysis of the biomechanical dynamics of AAAs and their rupture potential.

\section{CRediT authorship contribution statement}
Mario de Lucio: Writing – original draft, Software, Validation, Methodology, Investigation,  Conceptualization. Jacobo Díaz: Conceptualization, Software, Writing – review \& editing, Supervision, Methodology, Funding acquisition. Alberto de Castro: Writing – review \& editing, Resources. Luis E. Romera: Writing – review \& editing, Funding acquisition, Supervision.

\section{Declaration of competing interest}
The authors have no competing interests to declare.

\section*{Acknowledgments}
This research has been partly supported by the Galician Government (\emph{Xunta de Galicia, Consellería de Educación, Ciencia, Universidades e Formación Profesional}) under grant agreement ED431C 2021/33 (\emph{Axudas para a consolidación e estruturación de unidades de investigación competitivas e outras accións de fomento nas universidades do SUG - GRC}).







\nocite{*} 







\end{document}